\documentclass[preprint]{aastex61}

\received{}
\revised{}
\accepted{}
\submitjournal{ApJ}

\shorttitle{Li-enhanced metal-poor stars}
\shortauthors{Li et al.}

\begin{document}

\title{Enormous Li-enhancement preceding red giant phases in low-mass stars in the Milky Way halo\footnote{This work is based on data collected at the Subaru Telescope, which is operated by the National Astronomical Observatory of Japan.}}

\correspondingauthor{Wako Aoki}
\email{aoki.wako@nao.ac.jp}

\author[0000-0002-0786-7307]{Haining Li}
\affil{Key Lab of Optical Astronomy, National Astronomical Observatories, Chinese Academy of Sciences, Beijing 100012, China}

\author{Wako Aoki}
\affiliation{National Astronomical Observatory of Japan, Mitaka, Tokyo 181-8588, Japan}
\affiliation{Department of Astronomical Science, School of Physical Sciences, SOKENDAI (The Graduate University for Advanced Studies), Mitaka, Tokyo 181-8588, Japan}

\author{Tadafumi Matsuno}
\affiliation{National Astronomical Observatory of Japan, Mitaka, Tokyo 181-8588, Japan}
\affiliation{Department of Astronomical Science, School of Physical Sciences, SOKENDAI (The Graduate University for Advanced Studies), Mitaka, Tokyo 181-8588, Japan}

\author{Yerra Bharat Kumar}
\affil{Key Lab of Optical Astronomy, National Astronomical Observatories, Chinese Academy of Sciences, Beijing 100012, China}

\author{Jianrong Shi}
\affil{Key Lab of Optical Astronomy, National Astronomical Observatories, Chinese Academy of Sciences, Beijing 100012, China}

\author{Takuma Suda}
\affil{Research Center for the Early Universe, Graduate School of Science, University of Tokyo, Hongo, Tokyo 113-0033, Japan}

\author{Gang Zhao}
\affil{Key Lab of Optical Astronomy, National Astronomical Observatories, Chinese Academy of Sciences, Beijing 100012, China}



\begin{abstract}

Li abundances in the bulk of low-mass metal-poor stars are well
reproduced by stellar evolution models adopting a constant initial
abundance. However, a small number of stars have exceptionally high Li
abundances, for which no convincing models have been established. We
report on the discovery of 12 very metal-poor stars that have large
excesses of Li, including an object having more than 100 times higher
Li abundance than the values found in usual objects, which is the the
largest excess in metal-poor stars known to date. The sample is
distributed over a wide range of evolutionary stages, including five
unevolved stars, showing no clear abundance anomaly in other
elements. The results indicate the existence of an efficient process
to enrich Li in a small fraction of low-mass stars at the
main-sequence or subgiant phase.  The wide distribution of Li-rich
stars along the red giant branch could be explained by dilution of
surface Li by mixing that occurs when the stars evolve into red
giants. Our study narrows down the problem to be solved to understand
the origins of Li-excess found in low-mass stars, suggesting the
presence of unknown process that affects the surface abundances
preceding red giant phases.

\end{abstract}

\keywords{stars: abundances --- stars: evolution --- stars: low-mass --- stars: Population II --- nuclear reactions, nucleosynthesis, abundances}



\section{Introduction} \label{sec:intro}

Lithium (Li) abundances in the bulk of unevolved low-mass stars with
low metallicity are almost constant, which is recognized as a result
of the Big Bang nucleosynthesis \citep{spite82}, although a problem
that the value is systematically lower than that expected from
standard models remains \citep[e.g., ][]{cyburt16}. Li is, on the
other hand, destroyed by nuclear reactions inside stars with
temperatures higher than 2.5 million K. The surface Li is diluted by
more than one order of magnitude by mixing with the layers in which Li
is already depleted when a star evolves into a red giant. This is the
feature found by systematic observations for globular cluster stars
\citep{lind09}, demonstrating the overall success of the structure and
evolution models for low-mass stars.

There is, however, a small fraction of low-mass stars that show
extremely high Li abundances \citep{kumar11}, including globular
cluster stars \citep{kraft99}, and field metal-poor stars
\citep{ruchti11}. Whereas the Li production by the so-called
Cameron-Fowler mechanism \citep{cameron71} in the hot-bottom burning
phase of AGB stars is identified as the source of Li-excess found in
highly evolved, luminous objects \citep{smith89}, no models have been
established to explain this phenomenon for less luminous stars, making
this problem a challenge to the theory of low-mass star evolution.

Scenarios proposed to explain the Li-excess in low-mass stars include
i) extra mixing in red giants, in particular at the so-called red
giant branch (RGB) bump \citep{sackmann99, charbonnel00}, ii)
accretion of rocky planets that contain Li \citep{ashwell05}, iii)
mass transfer from Li-enhanced AGB or highly evolved red giants. 
  Other Li production processes in a variety of sites and phenomena,
  e.g., novae \citep{jose98} and supernovae \citep{woosley90}, are
  included in models of the Galactic chemical evolution, no
  quantitative models that explain Li-rich objects by such processes
  have been established. To examine such possibilities, observational
studies of very metal-poor stars have advantages. One is that
metal-poor stars are generally low-mass objects with long lifetimes,
in contrast to metal-rich stars with Li-excess that include
intermediate-mass objects \citep{kumar11}, which could make the
discussion complicated. Another advantage is that the ‘normal’
Li abundance in main-sequence and subgiant phases is well defined for
metal-poor stars, which is the so-called Spite plateau value
\citep{spite82, ryan99}. Moreover, chemical composition of metal-poor
stars is sensitive to the additional contribution of nucleosynthesis
that caused Li-enhancement. Correlation of Li-excess and abundance
anomaly of other elements, if any, could be a strong constraint on
proposed scenarios.

Observational studies of very metal-poor stars with Li-excess are,
however, still quite limited because of the low frequency of these
objects. \citet{ruchti11} report six Li-rich giants with
[Fe/H]$<-1.7$. \citet{roederer08} report a Li-enhanced star, HKII
17435-00532, around the RGB bump, showing enhancements of both r- and
s-process elements. Another Li-enhanced giant, CS 22893-010, is
reported by \citet{roederer14}. On the other hand, five Li-enhanced
stars (not in the AGB phase) have been reported in very metal-poor
star clusters. In NGC 6397, a very Li-rich star, up to $A({\rm
  Li})\sim 4$, located near the main sequence turn-off point was found
by \citet{koch11}. Other four stars are red giants found in M~68,
NGC~5053, and NGC~5897 \citep{kirby16}.


Here we report on discoveries of very metal-poor stars with large
enhancement of Li in the Milky Way halo, and discuss their chemical
properties and evolutionary stages. 

\section{Observations} \label{sec:obs}

Li-rich stars have been found in our program to study metal-poor stars
by the low-resolution spectroscopy with the Large sky Area
Multi-Object fiber Spectroscopic Telescope \citep[LAMOST:][]{cui12, zhao12} and
high-resolution follow-up observation with the Subaru Telescope.
Candidates of metal-poor stars are selected based on the results of a
pipeline analysis and visual inspection of the spectra \citep{li15}. Li-enhanced
objects are then selected by visual inspection for the wavelength
region around the Li 6708~{\AA} line. The sample consists of both red
giants and unevolved stars, i.e. main-sequence turn-off stars and
subgiants.

High-resolution spectroscopy was carried out for the candidates using
the Subaru Telescope High Dispersion Spectrograph \citep[HDS:
][]{noguchi02}. The spectra cover 4030--6760~{\AA} with $R=$60,000 and
signal-to-noise ratios from 50 to 200, sufficient to determine
abundance ratios of elements including Li. Table 1 lists the 12 stars
that are observed with sufficient quality and turn out to be extremely
Li-rich according to our analysis, as described below. Here, we select
objects that have $A({\rm Li})>2.0$ for red giants and $A({\rm Li})>
3.0$ for main-sequence turn-off stars and subgiants, which are at
least about one order of magnitude higher than the Li abundances of typical
metal-poor stars at similar evolutionary status (see below)\footnote{Abundance data are presented by $A({\rm Li})=\log(N_{\rm Li}/N_{\rm H}) + 12$ for Li, and [X/Y]$=\log(N_{\rm X}/N_{\rm Y}) - \log(N_{\rm X}/N_{\rm Y})_{\odot}$ for elements X and Y.}. For
comparison purposes, high-resolution spectra of bright stars for which
Li abundances are measured by previous studies are analyzed with the
same technique (the four objects from the bottom in Table 1). Spectra
of Li lines are shown in Figure 1. These were found in a sample of
about 950 candidates of very metal-poor stars, indicating that the
fraction of Li-rich objects is on the order of 1\% in metal-poor stars, which is similar to the estimate by \citet{ruchti11}.

High-resolution spectra are obtained at two or more epochs for most of
the targets to investigate variations of radial velocities. A time
variation of the radial velocity is detected for only one object
(J1314+3741). Hence, no signature of high binary frequency has been
found for Li-rich stars. Spectral line widths are also investigated
for the high-resolution spectra. No significant excess of line widths
due to stellar rotation is detected in our sample, except for
J0554+5235 of which spectral lines suggest the rotation as rapid as 7
km~s$^{-1}$.  The strong Li absorption features show no detectable
changes between different exposures.

 Ten stars in our sample are included in the
  photmetry data of Catalina
  Survey\footnote{http://nesssi.cacr.caltech.edu/DataRelease/}. Except for
  J0302+1356, which shows a variation of 0.5 magnitude in the $V$-band,
  our objects show no variation within the measurement errors ($\sim 0.05$
  magnitude) in the past decade. Namely, no clear anomaly is found in
  photometry monitoring for most of our sample.

\section{Analysis and results}

Stellar parameters, i.e., effective temperature ($T_{\rm eff}$),
surface gravity ($g$), metallicity and micro-turbulent velocity, are
determined by standard abundance analysis for spectral lines of
neutral and singly ionized Fe, adopting the line list of \ion{Fe}{1}
and \ion{Fe}{2} of \citet{aoki13} supplemented by data of
\citet{obrian91} and \citet{fuhr88}. The results are
given in Table 1. We also estimate $T_{\rm eff}$ from colors using
temperature scales of \citet{alonso96, alonso99},
\citet{casagrande10}, and \citet{ramirez05}. The $(V-K)_{0}$ values
derived from the APASS \citep{henden16} and 2MASS \citep{cutri03}
photometry data and reddening corrections from the dust map
\citep{schlafly11} available in the NASA/IPAC website are given in the
table.  For the five objects with $T_{\rm eff}$ higher than 5500~K, in
which the Balmer line profiles are sensitive to the changes of $T_{\rm
  eff}$ and $g$, these parameters are also estimated by the profile
fitting for the Balmer lines following the procedure of
\citet{barklem02} and \citet{matsuno17}. The results are given in
Table 2. The $T_{\rm eff}$'s obtained by these methods well agree with
the values obtained by spectroscopic analysis of Fe lines.

The $\log g$ values of comparison stars are also estimated based on
the parallax provided by Tycho-Gaia Astrometric Solution \citep[TGAS;
][]{lindegren16}. The $\log g$ values of the two warm comparison
stars (HD~84937 and HD~140283) agree with those from the Balmer line
analysis, whereas they are larger than those obtained by the
spectroscopic analysis by 0.3--0.4~dex. Similar discrepancy is found
between the values from spectroscopic analysis and from Balmer line
analysis for the five warm Li-rich stars. This could be due to the
non-NLTE (NLTE) effects in the analysis of Fe lines. The NLTE effect
on $\log g$ estimates is reported to be 0.2--0.5~dex by
\citet{lind12} for the evolutionary status
and metallicity ranges of our sample.

The typical uncertainty of the $T_{\rm eff}$, $\log g$, microturbulent
velocity, and [Fe/H] for very metal-poor stars are 150~K, 0.3~dex,
0.3~km~s$^{-1}$, and 0.3~dex, respectively.  We adopt the $T_{\rm eff}$ 
and $\log g$ obtained in the LTE analysis of Fe lines in the
following abundance measurements. The changes of $\log g$ values by
0.3~dex, suspected from the estimates by other methods, do not
significantly affect the results of elemental abundances.

A standard abundance analysis is carried out using the local
thermodynamic equilibrium (LTE) model stellar photospheres
\citep{castelli97}. The Li subordinate line at 6103.6~{\AA} is
detected in nine out of the 12 objects in our sample (Table 2), in
which the Li resonance line at 6707.8~{\AA} is highly saturated
(Figure 2). The LTE and NLTE Li abundances are determined from the
subordinate line for these objects. The abundances are also determined
from the resonance line for all targets including comparison
stars. The NLTE analysis is based on the method from
\citet{shi07}. Results are given in Table 2. The LTE abundance derived
from the subordinate line well agrees with the NLTE result, while the
resonance line would require NLTE analysis for desaturation. The NLTE
abundances from the two Li lines agree quite well. We adopt the
average of the NLTE results from the two lines as the final Li
abundance whenever available (Table 2).



Abundances of other key elements, C, Na, Mg and Ba, are determined by
standard analysis for equivalent widths of atomic lines and by the
spectrum synthesis method for the CH molecular band at 4315~{\AA} and
for species with only one atomic line. The line lists of
\citet{aoki13} for atoms and \citet{masseron14} for CH molecule
are adopted. The results are given in Table 2. The Na abundances are
determined from the subordinate lines at 5682 and 5688~{\AA} for 10 out of
the 12 objects, which are not sensitive to NLTE effects. For
J0554+5235 and J1455+1251, for which subordinate lines are not
detected, the Na abundances are estimated from the resonance lines at
5890 and 5896~{\AA}. For metal-poor stars with similar parameters, NLTE 
corrections are estimated to be as large as
$-0.5$~dex \citep{andrievsky07}. Excluding the Na abundances for these
two stars, the typical uncertainties of elemental abundance ratios are
0.12~dex, 0.31~dex, 0.13~dex, 0.15~dex and 0.18~dex for Li, C, Na, Mg
and Ba, respectively. 

The abundance ratios of C, Mg, and Ba determined for the Li-rich stars are typical values in very
metal-poor stars. Na is overabundant in some objects, but its abundance is
within the scatter found for very metal-poor stars with normal Li abundances. Hence, no signature
of peculiarity of abundance ratios for elements other than Li is found
in our sample. The exception is the most metal-poor object J0705+2552
that shows excesses of C, Na, and Mg. Such objects are, however, often
found in extremely metal-poor stars with no Li-excess
\citep{aoki02}. We exclude this object from the following discussion
and will report on the detailed properties separately.

\section{Discussion}

Figure 3 shows the stellar parameters and Li abundances of the Li-rich
objects studied in the present work. The luminosity ($L$) in the left
panel is estimated from $T_{\rm eff}$ and $g$ assuming a stellar mass
of 0.8~M$_{\odot}$. For this purpose, we adopt a NLTE correction for
$\log g$ estimated by \citet{lind12} mentioned in \S 3. Li-rich
metal-poor stars appear over a wide range of evolutionary stages in
low-mass stars. In particular, our sample includes five subgiants,
which should not have experienced significant mixing that occurs when
a star evolves into a red giant (the first dredge-up).  Only one warm
Li-rich object was previously known in the globular cluster NGC 6397
\citep{koch11}. A few other objects with less significant excess of Li
have been found near the bottom of the red giant branch in globular
clusters \citep{kirby16}. Our study reveals that Li-excess in low-mass
objects below the RGB bump, having luminosity lower than 100
L$_{\odot}$, is not a unique phenomenon in clusters, but is also found
in field metal-poor stars.
 
The metallicity range in which Li-excess appears is also quite wide,
although all our targets are very metal-poor ([Fe/H]$<-1.7$). 

The highest Li abundance, $A({\rm Li})=4.5$, is found in the subgiant
J0741+2132. The Li excesses in other subgiants and red giants below
the RGB bump are not as significant as in J0741+2132. The surface Li
is diluted by the first dredge-up that mixes the Li-containing layer 
($\sim$0.03~$M_{\odot}$) with Li-free layer of the envelope ($\sim
  0.4$~M$_{\odot}$). An example of model predictions is
depicted in the right panel of
Figure 3. If a very large Li excess up to $A({\rm Li})=4.5$ is assumed
as an initial value, the Li excess of stars with $A({\rm Li})\sim 3$
below the RGB bump can be explained by the dilution due to the first
dredge-up. This is a natural explanation for the presence of Li-rich
objects at any evolutionary status, which was speculated by a previous
study for globular cluster stars \citep{kirby16}. This hypothesis now
has observational support for the first time with our systematic
search for Li-rich very metal-poor stars. The probability of finding
red giants with $T_{\rm eff}\sim 5000$~K is several times higher than
that of finding subgiants with $T_{\rm eff}\sim 6000$~K in a
magnitude-limited sample, according to stellar evolution models
\citep[e.g.,][]{kim02}. This is because of the higher luminosity of
red giants, even though the time scale of their evolution is
shorter. Hence, if the frequency of Li-rich stars is similar among
both subgiants and red giants, it is no wonder that Li-rich giants
have been observed before our discovery of Li-rich subgiants. We note
that there are some very Li-rich objects with higher luminosity, some
of which are quoted to be AGB stars in literature. There could be
another origin of Li-excess for these highly evolved stars.

If the above interpretation is correct, the essential problem is the
source of the extremely large Li-excess in main-sequence turn-off
stars or subgiants like J0741+2132. Since no significant internal
mixing (with short timescale) is expected in such unevolved stars, at
least within the framework of standard stellar evolution theory, Li
production by the Cameron-Fowler mechanism in a single star is very
unlikely. Hence, interaction with other objects would be necessary.

Engulfment of planets having a high Li abundance (e.g. rocky planets)
is proposed as a possible origin of Li excess \citep{siess99}. This
is, however, also unlikely as efficient planet formation is not
expected in very metal-poor stars. Moreover, the amount of Li
contained in J0741+2132 is too high to be explained by this
scenario. Assuming 0.03 M$_{\odot}$ for the surface convective layer
with $A({\rm Li})=4.5$, the Li mass contained in the layer is about
$10^{-8.2}$ M$_{\odot}$. To provide this amount of Li from material
with primordial abundance ($A({\rm Li})\sim 2.2$), the total mass of
original material required is larger than 1~M$_{\odot}$.  Formation of
planets in which Li is concentrated from material with such large
mass, and subsequent accretion to the stellar surface, is highly
improbable.

A remaining scenario is accretion of matter affected by Li production
in a companion star in the AGB or highly evolved red giant phase. The
companion could be an unseen white dwarf at present. There is,
however, no signature of high binary frequency in our sample from
radial velocity monitoring nor from photometry. Moreover, there
is no signature of excess in carbon and neutron-capture elements,
which are usually regarded as a signature of mass transfer from an AGB
companion.

Recently, nova explosions have been identified as a promising source
of Li in the universe by the measurement of Li and $^{7}$Be in
nova spectra \citep{tajitsu15,
  izzo15}. No significant excesses of other elements have been found 
in such observations for novae. Accretion of
material ejected from a nova to a low-mass star is an attractive idea
to explain the Li-rich stars \citep{gratton89}. There is, however, no
model of mass accretion from a nova to a low-mass star at
present. Hence, this is still a speculation just from the abundance
properties.

Another scenario to be investigated is an extra mixing caused by
merging events with planets or other small mass objects, which might
induce the Cameron-Fowler mechanism even in stars before the RGB
bump \citep{denissenkov04}. Such merging events could result in high rotation velocity of
the surface of stars by additional angular momentum brought by the
small objects. No signature of rapid rotation is found in our sample. 
However, this does not exclude the above possibility, because the objects could
be sufficiently old so that the rotation has already become slow after
the merging event. Another possible signature of a merging event is
the increase in total stellar mass as suggested for the formation of
blue straggler stars. Accurate mass estimates for these objects by
high precision luminosity (i.e., distance) measurement or by stellar
seismology are highly desirable to examine this possibility.

\section{Summary and concluding remarks}

This work reports elemental abundances of 12 very metal-poor stars with
large excess of Li. This is the largest sample of such objects
covering a wide range of evolutionary stages. The existence of Li-rich
stars indicates that there are still unknown processes in low-mass
star evolution, even in the phase before they evolve into red giants,
the study of which is usually regarded as a well-established field in
astrophysics. Solving the mystery of their origin will provide new insight into
the structure and evolution of low-mass stars, which could propagate
in the studies of planetary systems and of Galaxy formation based on
observations of low-mass stars.


\acknowledgments


Funding for LAMOST (www.lamost.org) has been provided by the Chinese
NDRC. LAMOST is operated and managed by the National Astronomical
Observatories, CAS. This work was supported by NSFC grants
No. 11573032, 11233004, 11390371, 11473033, and 11550110492, JSPS
KAKENHI Grant Numbers 16H02168, 16K05287 and 15HP7004, and JSPS - CAS
Joint Research Program.

\vspace{5mm}
\facilities{LAMOST, Subaru}

\software{IRAF}  

\begin{longrotatetable}
\begin{deluxetable*}{llccccccccccccc}
\tablecaption{Effective temperature and surface gravity estimated by different methods \label{tab:param}}
\tablecolumns{15}
\tablenum{1}
\tablewidth{0pt}
\tablehead{
\colhead{ID} &\colhead{Object} & \colhead{$(V-K)_{0}$} & &
\colhead{$T_{\rm eff}$ (K)} & \colhead{$\log g$} & &
\multicolumn{3}{c}{$T_{\rm eff}$ (K)} & & 
\colhead{$T_{\rm eff}$ (K)} & \colhead{$\log g$} & & \colhead{$\log g$} \\
\cline{5-6} 
\cline{8-10} 
\cline{12-13} 
\cline{15-15} 
\colhead{} & & & &
\multicolumn{2}{c}{spectroscopic} & &
\colhead{A96/99} & \colhead{C10} & \colhead{RM05} & &
\multicolumn{2}{c}{Balmer} & & \colhead{parallax} 
}
\startdata
J0302+1356 & LAMOST~J030209.33+135656.3 & 2.048 && 5206 & 2.30    && 5041  & \nodata & 4998  &&\nodata &\nodata && \nodata \\
J0554+5235 & LAMOST~J055408.54+523559.0 & 1.799 && 5638 & 1.80    && 5367  & \nodata & 5319  &&\nodata &\nodata &&\nodata \\
J0626+6032 & LAMOST~J062647.91+603254.0 & 1.467 && 5885 & 3.45    && 5876  & 5987  & 5844  && 5873 & 3.55 &&\nodata \\
J0705+2552 & LAMOST~J070542.30+255226.6 & 1.798 && 5269 & 2.50    && 5432  & \nodata & 5400  &&\nodata &\nodata &&\nodata \\
J0714+1600 & LAMOST~J071422.66+160042.5 & 1.895 && 5179 & 2.40    && 5243  & \nodata & 5197  &&\nodata &\nodata &&\nodata \\
J0741+2132 & LAMOST~J074102.07+213246.6 & 1.270 && 6142 & 3.65    && 6181  & 6365  & 6299  && 6150 & 3.91 &&\nodata \\
J0758+4703 & LAMOST~J075816.39+470343.3 & 1.498 && 6151 & 3.65    && 5760  & 5924  & 5786  && 6093 & 3.99 &&\nodata \\
J0852+2627 & LAMOST~J085208.07+262730.1 & 1.604 && 5872 & 3.55    && 5597  & 5750  & 5591  && 5713 & 3.74 &&\nodata \\
J1314+3741 & LAMOST~J131457.78+374110.7 & 1.564 && 5809 & 3.15    && 5752  & 5825  & 5715  && 5699 & 3.83 &&\nodata \\
J1414+0016 & LAMOST~J141412.27+001618.7 & 2.133 && 4882 & 1.55    && 4955  & \nodata & 4936  &&\nodata &\nodata &&\nodata \\
J1455+1251 & LAMOST~J145500.04+125106.2 & 2.360 && 4670 & 1.05    && 4708  & \nodata & 4713  &&\nodata &\nodata &&\nodata \\
J2146+2732 & LAMOST~J214610.13+273200.8 & 1.916 && 5243 & 2.75    && 5205  & \nodata & 5153  &&\nodata &\nodata &&\nodata \\
HD84937    & HD~84937                   & 1.262 && 6263 & 3.75    && 6195  & 6382  & 6318  && 6208 & 4.21 && 4.07  \\
HD140283   & HD~140283                  & 1.624 && 5647 & 3.10    && 5590  & 5723  & 5562  && 5596 & 3.65 && 3.52  \\
HD186478   & HD~186478                  & 2.504 && 4648 & 1.15    && 4574  & \nodata & 4578  &&\nodata &\nodata && 1.18  \\
HD2796     & HD~2796                    & 2.174 && 4832 & 1.05    && 4906  & \nodata & 4888  &&\nodata &\nodata && 1.78  \\
\enddata
\tablecomments{$T_{\rm eff}$ and $\log g$ obtained by spectroscopic analysis are adopted for abundance analysis. References -- A96/99: \citet{alonso96}, \citet{alonso99}; C10: \citet{casagrande10}; RM05: \citet{ramirez05}}
\end{deluxetable*}
\end{longrotatetable}

\begin{deluxetable*}{lcccccccccccccc}[b!]
\tablecaption{Elemental abundances \label{tab:res}}
\tablecolumns{15}
\tablenum{2}
\tablewidth{0pt}
\tablehead{
\colhead{ID} & &
\multicolumn{13}{c}{Abundances}  \\
\cline{3-15} 
\colhead{} & & 
\colhead{[Fe/H]} &&  
\multicolumn{2}{c}{$A({\rm Li})$ 6103.6~{\AA}} &&
\multicolumn{2}{c}{$A({\rm Li})$ 6707.8~{\AA}} &&
\colhead{$A({\rm Li})$} & \colhead{[C/Fe]} & \colhead{[Na/Fe]} & \colhead{[Mg/Fe]} & \colhead{[Ba/Fe]} \\
\cline{5-6} \cline{8-9}  \cline{11-11}
\colhead{} & &
 & &  
\colhead{LTE} & \colhead{NLTE} && 
\colhead{LTE} & \colhead{NLTE} && (adopted)
& & &
}
\startdata
J0302+1356 &&  -1.74  &&\nodata&\nodata&& 2.34 & 2.24  && 2.24 & -0.29  & -0.22  & 0.31  & 0.19  \\
J0554+5235 &&  -2.03  && 3.33 & 3.42 && 4.05 & 3.45  && 3.44 & 0.30  & 0.75* & 0.22  & 0.05  \\
J0626+6032 &&  -2.29  && 3.19 & 3.21 && 3.43 & 3.20  && 3.21 & 0.36  & 0.27  & 0.12  & 0.10  \\
J0705+2552 &&  -3.19  && 3.09 & 3.15 && 3.54 & 3.15  && 3.15 & 1.76  & 1.37  & 1.04  & 0.48  \\
J0714+1600 &&  -2.16  && 2.28 & 2.32 && 2.42 & 2.32  && 2.32 & 0.03  & -0.39  & 0.32  & -0.31  \\
J0741+2132 &&  -2.33  && 4.55 & 4.51 && 4.85 & 4.55  && 4.53 & 0.65  & 0.10  & 0.35  & -0.41  \\
J0758+4703 &&  -1.84  && 3.40 & 3.44 && 3.79 & 3.45  && 3.45 & 0.36  & 0.45  & 0.40  & 0.06  \\
J0852+2627 &&  -2.13  && 3.04 & 3.07 && 3.16 & 3.02  && 3.05 & 0.20  & 0.21  & 0.30  & 0.34  \\
J1314+3741 &&  -2.70  && 3.16 & 3.21 && 3.48 & 3.21  && 3.21 & 0.55  & 0.68  & 0.37  & -0.43  \\
J1414+0016 &&  -2.56  && \nodata & \nodata && 2.42 & 2.36  && 2.36 & 0.03  & -0.39  & 0.32  & -0.31  \\
J1455+1251 &&  -2.68  && \nodata & \nodata && 2.31 & 2.24  && 2.24 & -0.40  & 0.34* & 0.35  & -2.37  \\
J2146+2732 &&  -1.73  && 2.55 & 2.56 && 2.85 & 2.61  && 2.59 & -0.20  & 0.81  & 0.22  & 0.47  \\
HD84937    &&  -2.31  && \nodata & \nodata && 2.21 & 2.16 && 2.16 & 0.38 & -0.30  & 0.31 & -0.08 \\
HD140283   &&  -2.64  && \nodata & \nodata && 2.08 & 2.06 && 2.06 & 0.61 & -0.37  & 0.26 & -1.01 \\
HD186478   &&  -2.55  && \nodata & \nodata && $<-0.22$ & \nodata && $<-0.22$ & -0.33 & -0.32  & 0.52 & -0.08 \\
HD2796     &&  -2.49  && \nodata & \nodata && $<0.06$ & \nodata && $<0.06$ & -0.54 & -0.37  & 0.35 & -0.31 \\
\enddata
\tablecomments{Na abundances with asterisk are determined from the resonance lines whereas those of other stars are determined from the subordinate lines.}
\end{deluxetable*}


\begin{figure}[ht!]
\plotone{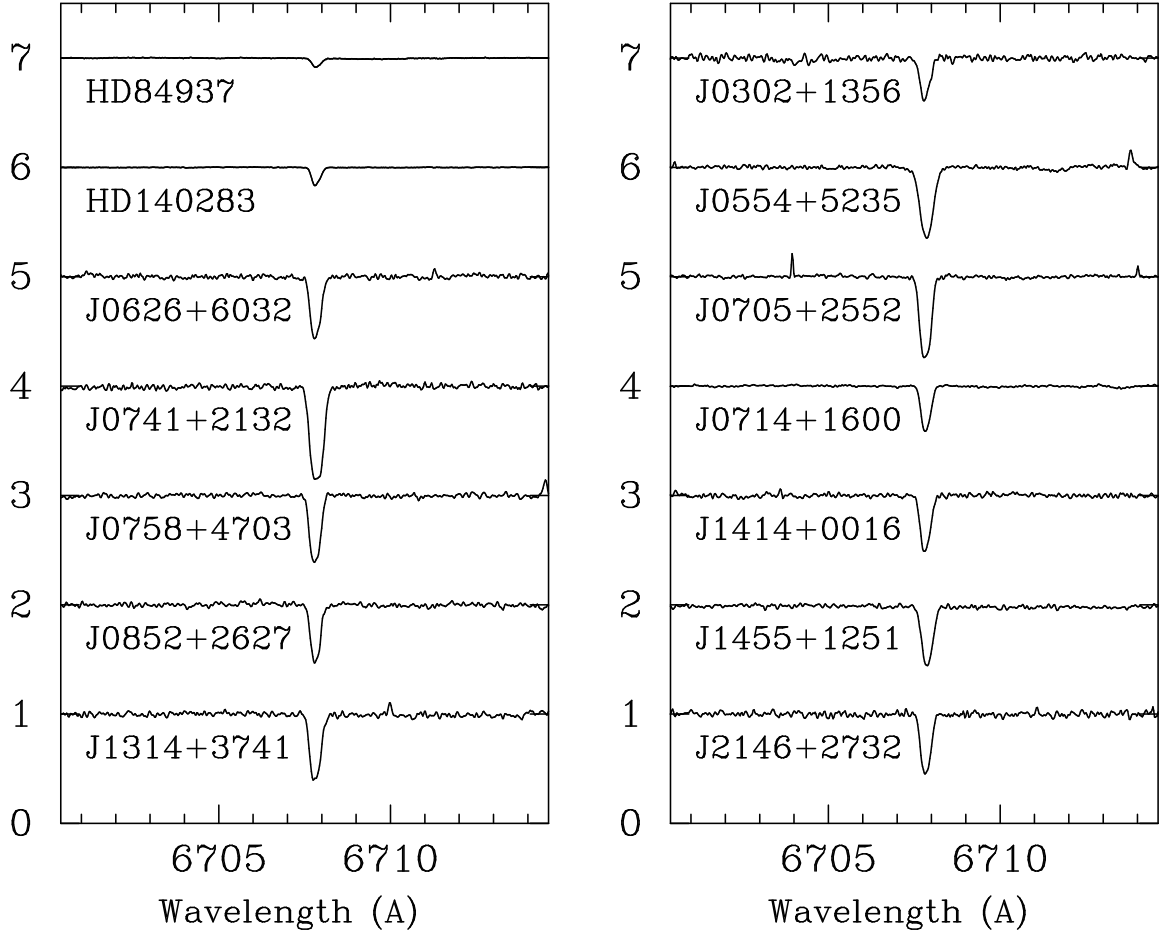}
\caption{
 Spectra around the Li line at 6707.8~{\AA} obtained with the
 Subaru/HDS. The spectra are normalized to the continuum level, and
 are vertically shifted for clarity. Object names are presented in the
 panels. The two objects from the top of the left panel are comparison
 stars that have normal Li abundances. 
%
}
\end{figure}

\begin{figure}[ht!]
\plotone{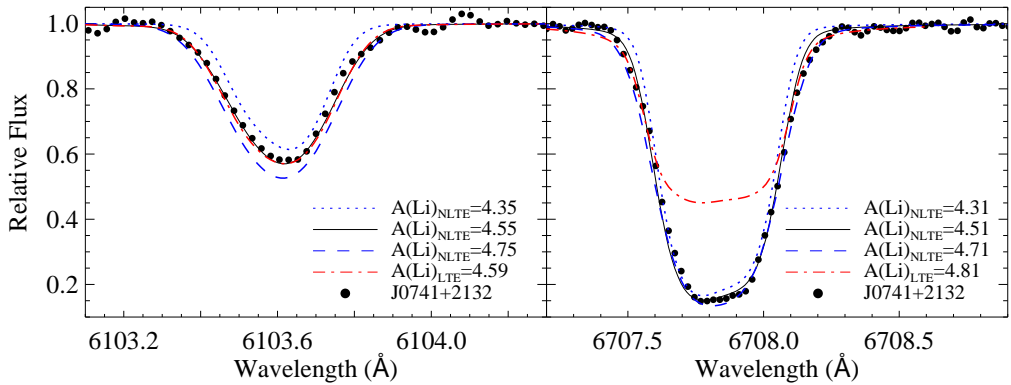}
\caption{ Spectra of the Li 6103.6~{\AA} and 6707.8~{\AA} lines of
  J0741+2132 (dots). Synthetic spectra with NLTE assumption for
  $A({\rm Li})=4.55 \pm 0.2$ dex are shown for the 6103.6~{\AA} line. A LTE
  spectrum also well reproduces the line profile by assuming
  $A({\rm Li})=4.59$ (dash-dotted line). By contrast, no LTE spectrum can
  explain the deep absorption profile of the 6707.8~{\AA} line. The NLTE
  spectrum for $A({\rm Li})=4.5$ reproduces the profile, though it is not as
  sensitive to the assumed abundance as the 6103.6~{\AA} line. 
}
\end{figure}

\begin{figure}[ht!]
\plotone{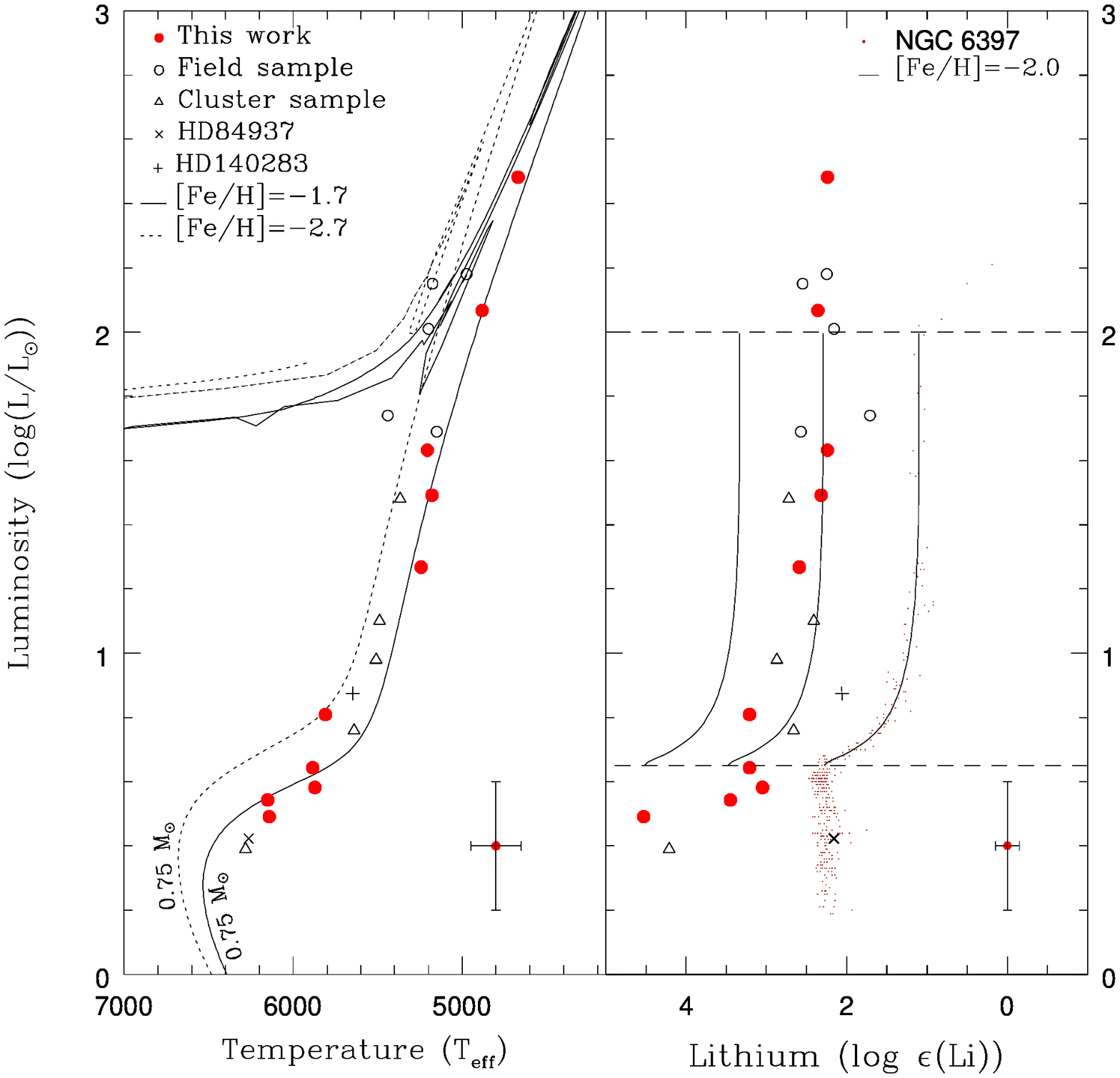}
\caption{Effective temperature and luminosity of Li-rich stars (left)
  and Li abundance and luminosity (right).  Our sample are shown by
  (red) filled circles, whereas data in literature with [Fe/H]$ <-1.7$
  are shown by open circles for field stars \citep{ruchti11,
    roederer08, roederer14} and open triangles for cluster stars
  \citep{koch11, kirby16}. The two comparison stars are shown by plus
  and cross symbols. The evolutionary tracks \citep{dotter16, choi16} for objects with
  0.75~M$_{\odot}$ for two cases of metallicity ([Fe/H]$=-1.7$ and
  $-2.7$). In the right panel, objects in the globular
  cluster NGC~6397 \citep{lind09} are shown by small dots. The solid
  lines show theoretical prediction of Li abundance as a function of
  luminosity by MIST24 \citep{choi16} for stars with 0.75~M$_{\odot}$ and
  [Fe/H]$=-2.0$ starting from the Spite plateau value $A({\rm
    Li})=2.2$ (the right line), and from $A$(Li)$=4.5$ (left) and
  3.5 (middle), the values close to those of J0741+2132 and J0758+4703,
  respectively. The dashed horizontal lines indicate the luminosity at
  which Li abundances show significant decreases in globular cluster
  stars \citep{lind09}.
}
\end{figure}




\end{document}